\documentclass[prl,twocolumn,amsmath,amssymb,superscriptaddress]{revtex4-2}

\usepackage{graphicx}
\usepackage{dcolumn}
\usepackage{bm}
\usepackage{color}
\usepackage{braket}

\begin{document}

\title{Charge Density Wave transient states induced by ultrashort hard X-ray pulses}

\author{D. Ghoneim}
\affiliation{Laboratoire de Physique des Solides, Université Paris-Saclay, CNRS, 91405 Orsay, France}
\affiliation{European X-Ray Free-Electron Laser Facility, D-22869 Schenefeld, Germany}
\author{E. Bellec}
\affiliation{Laboratoire de Physique des Solides, Université Paris-Saclay, CNRS, 91405 Orsay, France}
\author{L. Ortega}
\affiliation{Laboratoire de Physique des Solides, Université Paris-Saclay, CNRS, 91405 Orsay, France}
\author{V.L.R. Jacques}
\affiliation{Laboratoire de Physique des Solides, Université Paris-Saclay, CNRS, 91405 Orsay, France}
\author{A.A Sinchenko}
\affiliation{Laboratoire de Physique des Solides, Université Paris-Saclay, CNRS, 91405 Orsay, France}
\author{P. Wzietek}
\affiliation{Laboratoire de Physique des Solides, Université Paris-Saclay, CNRS, 91405 Orsay, France}
\author{U. Boesenberg}
\affiliation{European X-Ray Free-Electron Laser Facility, D-22869 Schenefeld, Germany}
\author{K. Kazarian}
\affiliation{European X-Ray Free-Electron Laser Facility, D-22869 Schenefeld, Germany}
\affiliation{Paul Scherrer Institute PSI Forschungsstrasse 111
5232 Villigen PSI Switzerland}
\author{J. Moeller}
\affiliation{European X-Ray Free-Electron Laser Facility, D-22869 Schenefeld, Germany}
\author{J.D. Moore}
\affiliation{European X-Ray Free-Electron Laser Facility, D-22869 Schenefeld, Germany}
\author{A. Rodriguez-Fernandez}
\affiliation{European X-Ray Free-Electron Laser Facility, D-22869 Schenefeld, Germany}
\author{R. Shayduk}
\affiliation{European X-Ray Free-Electron Laser Facility, D-22869 Schenefeld, Germany}
\author{A. Zozulya}
\affiliation{European X-Ray Free-Electron Laser Facility, D-22869 Schenefeld, Germany}
\author{A. Madsen}
\affiliation{European X-Ray Free-Electron Laser Facility, D-22869 Schenefeld, Germany}
\author{D. Le Bolloc'h}
\affiliation{Laboratoire de Physique des Solides, Université Paris-Saclay, CNRS, 91405 Orsay, France}
\date{\today}

\begin{abstract}

The interaction of quantum systems with ultrashort laser pulses may lead to transient states far from equilibrium. Among these, Charge Density Wave (CDW) systems offer very rich non-equilibrium physics when submitted to ultrashort optical excitations in the infrared regime. 
In this Letter, we show that CDW systems pumped by ultrashort x-ray pulses excite the system in very different way. The x-ray pump, 6000 times more energetic than a laser pulse, pumping and probing the whole bulk system simultaneously, drives CDW systems in a novel reversible out of equilibrium state. The successive CDW states observed in the NbSe$_3$ system with respect to the arrival of x-ray pulses follow a non-thermal kinetic path. It is characterized by a change in 2k$_F$ wavevector, correlation lengths and electronic gap, as well as a change of dimensionality, all of this takes place within a very long recovery time ranging from 450$ns$ to 0.01$s$. Several scenarios are proposed, including transient nucleation and recombination of charged topological objects. 
 
\end{abstract}


\maketitle

\section{Introduction}
X-ray based pump-probe experiments using ultrafast laser pumps have been extensively developed in recent years, providing a better understanding of the coupling between different degrees of freedom (charge, lattice, spin and orbitals) and leading to emergent phenomena far from thermodynamic equilibrium. CDW systems have been widely studied using this experimental approach because strong optical excitation triggers a wide range of responses, such as coherent collective excitations\cite{Huber2014, Moore2016}, decorrelated CDW phase and amplitude fluctuations\cite{Lee2012, Han2012}, non-thermal melting of the CDW phase\cite{Johnson2011, Erasmus2012, Mazin2022}, driven phase transition between two CDW phases\cite{Haupt2016,Laulhe2017, johnson2017}, transient CDW state\cite{Kogar2020, Zhou2019} or ultrafast melting of electronic order\cite{PhysRevLett.125.266402}.
In all cases, the  intensity of the satellite reflection abruptly drops in the femtosecond time scale after the laser excitation and the time needed to recover the CDW ground state is much longer, from picoseconds to $\mu$s.  The ultrafast optical pump induces disorder that appears in the form of domain walls, vortices, or dislocations, which reduce long-range order and slow the dynamics of recovery or may change the nature of the phase transition\cite{Yusupov2010,Mertelj2013, Han2012, Vogelgesang2017, Zong2019, Zhou2019}. 

The excitation of condensed matter by an ultrafast hard x-ray beam is an emerging field. Nonthermal melting events have been reported by x-ray pumping experiments in three semiconductors, silicon\cite{PhysRevLett.126.015703,Pardini2018NonthermalMeltingSi}, diamond\cite{PhysRevLett.126.117403} and Al$_2$O$_3$\cite{PhysRevLett.128.223203}. In these three studies, the x-ray dose used was much higher than the damage threshold, leading to a nonthermal, irreversible loss of crystal structure that occurs within a few tens of femtoseconds after the first x-ray pulse.

The physical properties of CDW systems are obviously quite different as the band gap is 10 times smaller than that of the semiconductors. The pulses intensity used in this study are much smaller than the threshold damaged, but still sufficient to induce a drastic change in the CDW ground state.

To illustrate the effect of x-ray pump in CDW systems, the archetype niobium triselenide inorganic compound NbSe$_3$ has been studied. It belongs to the class of transition metal trichalcogenides that crystallize in the monoclinic P2$_1$/m space group. The experiment was carried out below the first CDW transition, at T=80K, i.e. 64K below the transition temperature T$_c$=144K. It should be noted that the actual sample temperature was determined by  continuously measuring its resistance throughout the experiment. The Q(0 2 0) Bragg reflection and the Q$_s$(0 1+q$_{s}$ 0) satellite associated to the periodic lattice distortion, with the incommensurate wave vector q$_{s}$= $0.241$, have been observed. Rocking curves have been performed around the two reflections using x-ray pulses delivered by the XFEL source. 

\begin{figure*}[t]
\includegraphics[width=1.\textwidth]{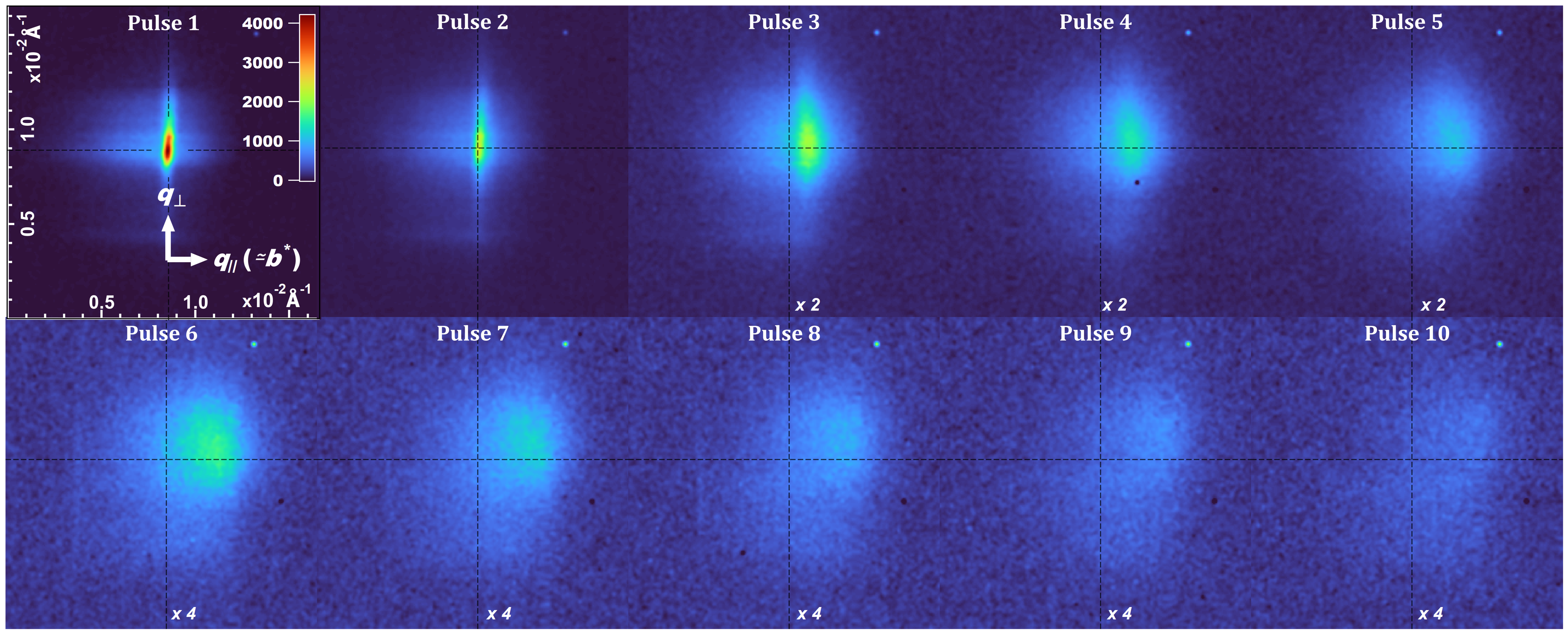}
\caption{Q$_{CDW}$ Q(0 1.24 0) satellite reflection associated to the CDW in NbSe$_3$ versus pulse number within a train of pulses. The process is reproducible from one train to the next. Each image corresponds of the sum over the rocking curve and have been averaged over 6000 trains. The intensity range has been rescaled for each pulse for better visibility (indicated in white on each image). The image size is 128*128 pixels of 200$\mu$m*200$\mu$m size (AGIPD detector) with 1.3 10$^{-4}$A$^{-1}$ per pixel in horizontal and 1.3 10$^{-4}$A$^{-1}$ in vertical. The intensity is multiply by either 2 or 4 for a better visibility as indicated on each diffraction pattern.}
\end{figure*}

The diffraction experiment was performed at the Materials Imaging and Dynamics (MID) instrument of the European x-ray Free Electron Laser (Eu-XFEL). The EuXFEL accelerator delivers x-ray pulses in bursts (named trains) with a repetition rate of 10Hz. For this study, up to 20 pulses per train with a repetition rate of 2.25 MHz were delivered, which corresponds to 444 $ns$ delay times between pulses. Each pulse has a duration of less than 50fs in the self-seeded mode with an energy bandwidth of approximately 1eV at 9$\pm 0.05$keV incident photon energy\cite{madsen2021,Liu2023Cascaded}. 

The high penetration of hard x-ray beams allows one to work in transmission geometry. 
The diffraction experiment has been performed across the 1$\mu m$ sample thickness, corresponding to a small absorption by the sample ($4\%)$.
Taking into account the low sample absorption, the 300*300$\mu m^2$ beam size which was vertically larger than the sample size (40$\mu$ m$(V) \times $ 500 $\mu$ m(H)) and attenuators used to reduce the intensity well below the damaged threshold, the deposited dose in the sample can be estimated at 0.8 mJ/cm$^2$.

A previous experiment using the LCLS XFEL source was performed to probe the same NbSe$_3$ system\cite{lebolloch_scienceadv2025}. The source delivered pulses shorter than 50 $fs$ at 120 Hz repetition rate—that is, one pulse every 8$ms$. In this frequency range, the satellite peaks remained stable from pulse to pulse. No transient effect induced by the ultrashort x-ray beam was measured. 

The present experiment performed at the Eu-XFEL, which generates pulses at a much higher frequency, leads to a radically different conclusion.
Although the satellite peak is stable from train to train, the satellite reflection Q$_s$ changes dramatically within trains (see Fig.1). A large shift of the 2k$_F$ wavevector is observed along the longitudinal direction versus the pulse number, as well as a continuous decrease in intensity and an increase in width. Although no change in the transverse direction is measured, the longitudinal component, on the other hand, is highly sensitive from pulse to pulse. Before disappearance after the 10th pulse, the longitudinal shift is $\delta q$=0.3 10$^{-3}\\A^{-1}$, corresponding to a large decrease of the CDW period $\lambda_{CDW}$ from 14.44 $\AA$ to 14.26 $\AA$.

The satellite peak completely disappears after the 10$^{th}$ pulse but the CDW melting is only transient. The satellite's peak returns to its initial state before the arrival of the first pulse of the next train, i.e. 0.1$s$ later. Most likely, the same effect occurred after each pulse during our previous experiment at LCLS. The repetition rate of the source was probably too small to allow us to observe reversible CDW melting.


\begin{figure}[t]
\includegraphics[scale = 0.37]{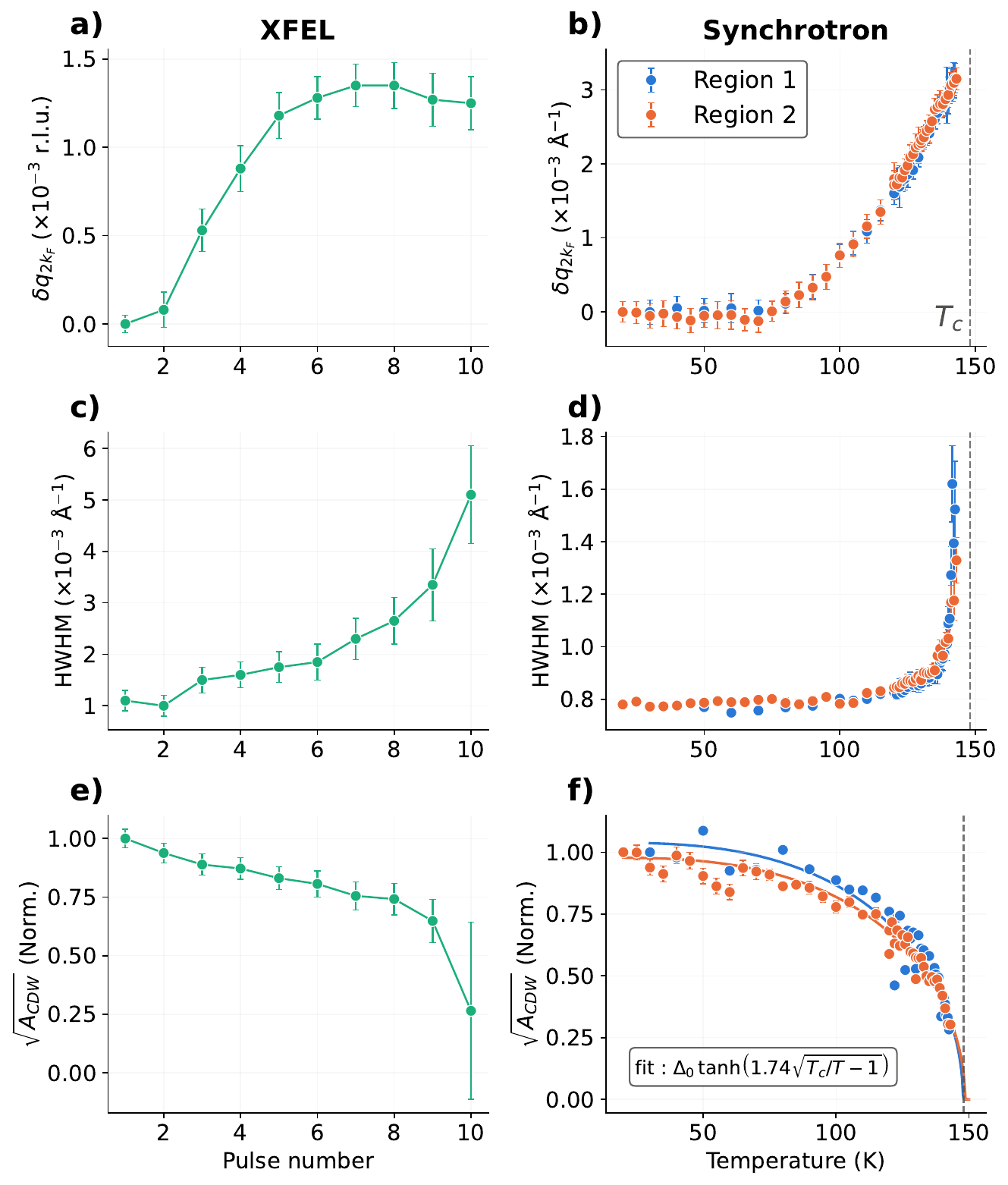}
\caption{(a, c, e) XFEL Pulse-number dependence of the CDW satellite shift $\delta q_{CDW}$ along the longitudinal direction ($\approx$b), the HWHM, and the square root of the CDW amplitude $\sqrt{\text{A}}_{CDW}$. (b, d, f) Temperature dependence (Synchrotron) of the same three quantities. on two regions of the same sample. The dashed vertical line marks the assumed critical temperature $T_c$ = 148 K; in (f), the solid curves are fits of $\sqrt{\text{A}}_{CDW}$(T) to the BCS-like mean-field form $\Delta_0\tanh\left(1.74\sqrt{T_c/T-1}\right)$}
\label{fig:compare}
\end{figure}

The transient melting of the CDW is not due a temperature increase induced by the X-ray beam itself.
The first evidence of this non-thermal phenomenon can be inferred from the behavior of the Bragg peak. The behavior of the Q(020) peak as a function of 20 pulses per train shows only a limited deviation. A slight change in profile is indeed observed as a function of pulse number, but the maximum of the rocking curve remains constant, given our resolution (see the Supplementary material). The heating effect induced by the x-ray beam on the host atomic lattice can be estimated to be less than 12K, well below the 64K required to undergo the transition temperature (T$_c$=144K).


\begin{figure}[t]
\includegraphics[scale = 0.48]{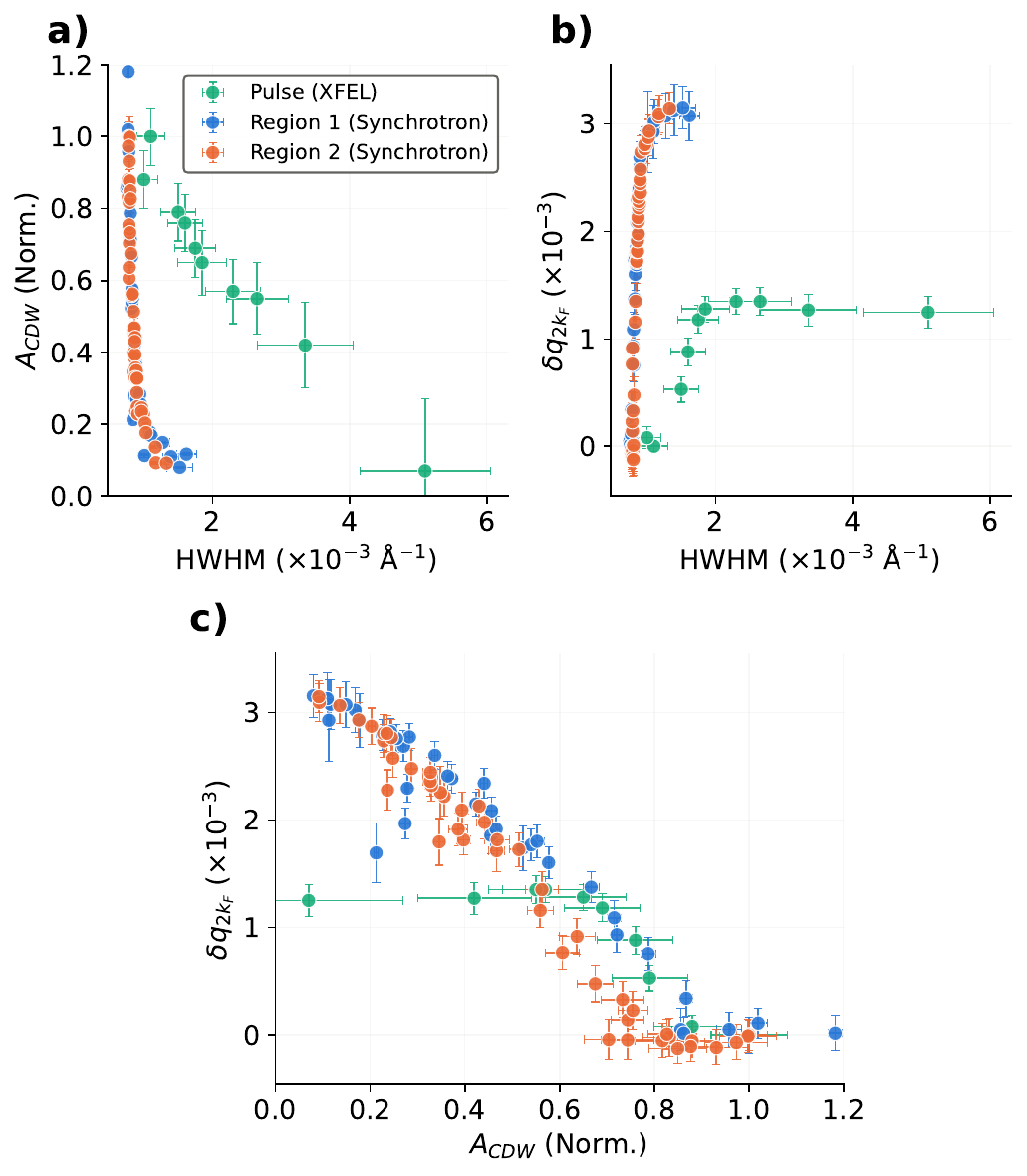}
\caption{a-c): Correlations between the three CDW order parameters shown in Fig.\ref{fig:compare}. a) A$_{CDW}$ versus the HWHM. b) The 2k$_F$ shift $\delta q_{CDW}$ versus the HWHM. c) $\delta q_{CDW}$ versus A$_{CDW}$. Error bars are propagated from each quantity’s fit uncertainty. As in Fig.\ref{fig:compare}, blue and orange dots are the synchrotron measurements as a function of temperature (on the two sample regions), and green dots are the XFEL data as a function of pump-pulse number.}
\label{fig:correlation}
\end{figure}

The previous observation regarding the host lattice is corroborated by the behavior of the satellite peak. Each CDW system is characterized by three key parameters, the correlations lengths $\xi$, the gap $\Delta$ and the CDW wavevector 2k$_F$. These three parameters depend on the topology of the Fermi surface, the electron-phonon coupling, and pinning by bulk impurities\cite{Fukuyama1978} and are closely linked to one another\cite{gruner2018density,Monceau2012}. These three parameters can be obtained from diffraction experiments through the 2k$_F$ value, the peak intensity ($I(q)\propto \Delta^2$) and the Half width Half Maximum of the satellite reflection ($\xi=1/\text{HWHM}$). 

The behavior of the satellite as a function of temperature was partially performed in NbSe$_3$ from a synchrotron source in \cite{mouddenPhysRevLett.65.223}. We completed the measurement at the BM02 beamline at the ESRF synchrotron source. In agreement with \cite{mouddenPhysRevLett.65.223}, the gap follows the usual BCS-like behavior ($\Delta(T)\propto\Delta_0\tanh(1.74\sqrt{T_c/T-1})$), the divergence of the inverse of the correlation length and the increase of the $2k_F$ value when approaching T$_c$ (see Figs 2f, 2d, 2a respectively).  A comparison of the behavior of the three parameters versus temperature and pulse number is shown in Fig.2.

In order to compare the two measurements —one taken at thermodynamic equilibrium versus temperature and the other versus pulse number at fixed temperature- we plot the gap as a function of 2k$_F$, the gap $\Delta$ versus correlation length $\xi$ and 2k$_F$ versus $\text{HWHM}$ (see Fig. 3). The huge difference between the two types of measurement highlights the non-thermal process induced by the ultrashort x-ray beam.

Another major difference between the 2k$_F$ behavior with temperature and the behavior observed here with pulses number is the loss of inhomogeneity. At equilibrium, the 2k$_F$ fluctuations in NbSe$_3$ above T$_c$ display  a large
anisotropy, leading to diffuse scattering elongated along the transverse directions, with the following ratio: $\xi_{b*}/\xi_{c*}$=3.5$\pm 0.5$ and $\xi_{b*}/\xi_{a*}$=27$\pm 0.2$ \cite{Moudden1993NbSe3q1}. In the transverse direction t*=c*+a*, this corresponds to $\xi_{b*}/\xi_{t*}$=12.5, far from the observation here. In the present case, the anisotropy ratio measured at the 2$^{nd}$ pulse ($\xi_t$/$\xi_{b*}$=5.2) is strongly reduced pulse after pulse until $\xi_t$/$\xi_{b*}$=2.2 at the 10$^{th}$ pulse (see Fig.1). The hard ultrashort x-ray beam reduces the inhomogeneity of CDW domains, moving the system away from a quasi-one-dimensional system.


X-ray pumping deeply changes the nature of the excited state by large 2k$_F$ shift, very slow relaxation, and loss of inhomogeneity.
This non-thermal behavior could be explained first by a decrease of electron-phonon coupling induced by the ultrashort x-ray pulse. The decreasing coupling would lead to a decreasing  band gap and thus a decreasing coherence lengths and loss of intensity because all three depend on the band gap ($\xi\propto \Delta^2$ and $I\propto \Delta^2$)\cite{Fukuyama1978,PhysRevB.50.8288}. However, a purely electronic phenomenon, very fast in principle, is hard to reconcile with the very slow relaxation times observed here and with the loss of inhomogeneity.

The change in 2k$_F$, as well as the increase in homogeneity as a function of pulse number, both associated with a slow recovery time, are more consistent with doped CDW systems.
Studies performed on NbSe$_3$ that substitute the transition-metal niobium element with tantalum\cite{PhysRevB.50.8288} or titanium impurities\cite{PhysRevB.59.15121} show a large decrease of the anisotropy ratio $\xi_{b*}$:$\xi_{a*}$:$\xi_{c*}$. In the case of NbSe$_3$ doped with Ti-1\%\cite{PhysRevB.59.15121}, the anisotropy ratio is 4:2:1 far from the large anisotropy present in the pure sample (27:8:1.7). Up to 40K above the transition temperature, the diffuse scattering becomes more and more asymmetric (see Fig.3 in \cite{mouddenPhysRevLett.65.223}). NbSe$_3$ doped with impurities also strongly alters the 2k$_F$ value with $\delta q/2k_F=-0.014/0.241=-5.8\%$ with Ti-1\%\cite{PhysRevB.59.15121}. The negative 2k$_F$ shift  with doping results in decreasing band filling due to the substitution of Nb$^{5+}$ by Ti$^{4+}$.

Following similar thought, the behavior of NbSe$_3$ submitted to ultrashort x-ray pulses could nucleate local CDW distortions in the bulk. Since the 2k$_F$ shift with pulse number is positive, these distortions must carry an excess of charges. The x-ray beam would therefore dope the system through nucleation of topological objects such as solitons. In that case, this corresponds to a delocalized 2$\pi$ phase shift along the chain axis, leading to an additional CDW periods, each of them carrying 2 electrons\cite{PhysRevB.94.201120}. The doping concentration can be estimated through the 2k$_F$ shift. In NbSe$_3$ indeed, two electrons partially fill four bands out of six, giving an average $\langle 2k_F\rangle=\frac14$\cite{canadell1990}. The presence of additional solitons leads to $(2+6\times2x)$ electrons spread over the 4 bands where $x$ is the soliton concentration. The excess of charges would lead to a positive shift 2k$_F$=$\frac14$+$\frac32$x, that is to a maximum density of solitons x=$\frac{2}{3}\delta q=\frac{2}{3}\times1.4\ 10^{-3}\approx$ 0.1\%.

In conclusion, the successive CDW states observed in NbSe$_3$ with respect to pulse number are characterized by a change in 2k$_F$ wavevector, correlation lengths and electronic gap, as well as a change of dimensionality, within a very long recovery time. The proposed scenario, which involves the creation of charged topological defects, appears to be consistent with observations. The exact origin of their formation under ultrashort x-ray pulses remains to be explained, although it should be based on a profound modification of the Se-Se covalent bonds present in NbSe$_3$.

\begin{acknowledgments}
The authors would like to thanks S. Ravy for very interesting discussions. We acknowledge European XFEL in Schenefeld, Germany, for provision of X-ray free-electron laser beamtime at MID SASE2 instrument under proposal number xxx and would like to thank the staff for their assistance. Data recorded for the experiment at the European XFEL are available at doi:10.22003/XFEL.EU-DATA-xxxxx-00.
This research was supported in part through the Maxwell computational resources operated at Deutsches Elektronen-Synchrotron DESY.

\end{acknowledgments}

\bibliography{biblio_MID2021}

\end{document}